\documentclass[sigplan,nonacm]{acmart}

\usepackage{enumitem}
\usepackage{graphicx}
\usepackage{multirow}
\usepackage{subcaption} 
\usepackage{lipsum}  
\usepackage{soul,listings,xcolor}
\usepackage{adjustbox}
\usepackage{tabularx} 
\usepackage{booktabs} 
\usepackage{graphicx} 
\usepackage{makecell}
\usepackage{pifont}
\newcommand{\xmark}{\ding{55}}

\DeclareRobustCommand{\hlgreen}[1]{{\sethlcolor{lime}\hl{#1}}}

\iftrue
\newcommand{\jaeyeon}[1]{{\color{red}{Jaeyeon: #1}}}
\newcommand{\saman}[1]{{\color{orange}{Saman: #1}}}
\newcommand{\joel}[1]{{\color{blue}{Joel: #1}}}
\newcommand{\willow}[1]{{\color{olive}{Willow: #1}}}
\else
\newcommand{\jaeyeon}[1]{}
\newcommand{\saman}[1]{}
\newcommand{\joel}[1]{}
\newcommand{\willow}[1]{}
\fi

\AtBeginDocument{%
  }

\setcopyright{none} 
\copyrightyear{2025}
\acmYear{2025}
\acmDOI{XXXXXXX.XXXXXXX}

\begin{document}



\title{Insum: Sparse GPU Kernels Simplified and Optimized with Indirect Einsums}


\author{Jaeyeon Won}
\affiliation{%
  \institution{Massachusetts Institute of Technology}
  \city{Cambridge}
  \country{USA}
}
\email{jaeyeon@csail.mit.edu}

\author{Willow Ahrens}
\affiliation{%
  \institution{Georgia Institute of Technology}
  \city{Atlanta}
  \country{USA}
}
\email{ahrens@gatech.edu}

\author{Joel S. Emer}
\affiliation{%
  \institution{Massachusetts Institute of Technology}
  \city{Cambridge}
  \country{USA}
}
\affiliation{%
  \institution{NVIDIA}
  \city{Westford}
  \country{USA}
}
\email{emer@csail.mit.edu}

\author{Saman Amarasinghe}
\affiliation{%
  \institution{Massachusetts Institute of Technology}
  \city{Cambridge}
  \country{USA}
}
\email{saman@csail.mit.edu}




\begin{abstract}



Programming high-performance sparse GPU kernels is notoriously difficult, requiring both substantial effort and deep expertise. Sparse compilers aim to simplify this process, but existing systems fall short in two key ways. First, they are primarily designed for CPUs and rarely produce high-performance GPU code. Second, when computations involve both sparse and dense regions, these compilers often fail to optimize the dense portions effectively. In this paper, we propose a new approach for expressing sparse computations. We start from format-agnostic Einsums over sparse tensors and rewrite them into format-conscious indirect Einsums, which explicitly encode format information by mapping sparse data and metadata onto dense tensor operations through indirect indexing. To execute indirect Einsums, we introduce the \textbf{Insum} compiler, which generates efficient GPU code for these Einsums by lowering to the PyTorch compiler, extended to better support Tensor Core–enabled indirect Einsums. We also present two fixed-length sparse formats, \textbf{GroupCOO} and \textbf{BlockGroupCOO}, designed to fit naturally with indirect Einsums. Our approach achieves $1.14\times$–$3.81\times$ speedups across a range of sparse GPU applications while reducing lines of code by $202\times$–$4491\times$ compared to hand-written implementations.




\end{abstract}



\maketitle

\section{Introduction}


Sparse computations arise naturally in many real-world applications, from machine learning models with graphs~\cite{torchgeometric,dgl} to scientific simulations~\cite{spsolve1,spsolve2}. In these scenarios, storing all values including many zeros leads to inefficient implementation, both in memory and computation.  To address this, developers rely on sparse formats that help avoid storing zeros and performing ineffectual computations ($a*0=0$).

At the same time, GPUs have become the hardware backbone of high-performance computing due to their massive parallelism. As demand grows for fast processing, there's a strong incentive to accelerate sparse workloads on GPUs.

Unfortunately, implementing efficient sparse kernels on GPUs is notoriously difficult. Unlike dense computations, sparse operations require indirect memory accesses, irregular data dependencies, and careful handling of load balancing. Developers must select appropriate sparse formats for their use case and manage a wide range of GPU-specific optimizations, including shared memory utilization, load balancing, vectorization, avoiding bank conflicts, exploiting tensor cores, and memory coalescing. These challenges significantly complicate the development of sparse GPU kernels.


These complexities often lead to specialized, hand-optimized GPU implementations, which are not only time-consuming to write but also difficult to maintain and generalize. The complexity of this problem is evident from the continuous influx of research papers introducing new implementations for sparse GPU workloads. For example, as shown in Table~\ref{tab:comparison}, TorchSparse~\cite{torchsparse1} is a state-of-the-art GPU sparse convolution library with over 4,000 lines of CUDA (NVIDIA’s low-level GPU programming language). Its implementation is so sophisticated that its underlying strategies have led to at least five separate research papers~\cite{torchsparse1,torchsparse2,torchsparse3,torchsparse4,torchsparse5}, each proposing novel techniques to address specific performance challenges.



\begin{table}[t]
\centering
\resizebox{0.48\textwidth}{!}{%
\begin{tabular}{|l|c|c|c|c|}
\hline
\textbf{Name} & 
\makecell{\textbf{Struct.} \\ \textbf{SpMM}} & 
\makecell{\textbf{Unstruct.} \\ \textbf{SpMM}} & 
\makecell{\textbf{Equivariant} \\ \textbf{Tensor Prod.}} & 
\makecell{\textbf{Sparse} \\ \textbf{Conv.}} \\
\hline
\texttt{TorchBSR\cite{quansight2023bsr}} & 202 LoC & \textcolor{red}{\xmark} & \textcolor{red}{\xmark} & \textcolor{red}{\xmark} \\
\texttt{Sputnik\cite{sputnik}} & \textcolor{red}{\xmark} & 1918 LoC & \textcolor{red}{\xmark} & \textcolor{red}{\xmark} \\
\texttt{e3nn\cite{e3nn}} & \textcolor{red}{\xmark} & \textcolor{red}{\xmark} & 225 LoC  & \textcolor{red}{\xmark} \\
\texttt{TorchSparse\cite{torchsparse1}} & \textcolor{red}{\xmark} & \textcolor{red}{\xmark} & \textcolor{red}{\xmark} & 4491 LoC \\
\Xhline{1pt}
Ours & 1 LoC & 1 LoC & 1 LoC & 1 LoC \\
\Xhline{1pt}
\textbf{LoC Saving} & 
\textcolor{green!60!black}{\textbf{202$\times$}} & 
\textcolor{green!60!black}{\textbf{1918$\times$}} & 
\textcolor{green!60!black}{\textbf{225$\times$}} & 
\textcolor{green!60!black}{\textbf{4491$\times$}} \\
\textbf{Speedup} & 
\textcolor{green!60!black}{\textbf{1.95$\times$}} & 
\textcolor{green!60!black}{\textbf{1.20$\times$}} & 
\textcolor{green!60!black}{\textbf{3.81$\times$}} & 
\textcolor{green!60!black}{\textbf{1.14$\times$}} \\
\hline
\end{tabular}
}
\caption{Comparison of different libraries across applications. While each library is tailored to a specific application, our compiler-based approach supports all of them, achieving better speedups with significantly less code.}
\label{tab:comparison}
\end{table}

Sparse tensor compilers have emerged to simplify the development of sparse kernels~\cite{kjolstad_tensor_2017,finch,mlirsparse,ye2023sparsetir,tacoucf}. These systems allow developers to express computations as if operating on dense tensors, while automatically generating optimized implementations for sparse formats. Typically, they achieve this separation through two key abstractions: (1) a format-agnostic Einsum~\cite{einstein_grundlage_1916} that specifies the computation (e.g., $C_i = A_i * B_i$ for elementwise multiplication), and (2) an explicit specification of the sparse storage format (e.g., \texttt{A} stored in a sparse format such as COO or CSR~\cite{chou2018format}).

Sparse compilers have largely focused on sparse-sparse kernels—operations where multiple sparse inputs interact. These kernels require identifying nonzero values at overlapping (intersection) or combined (union) coordinates, which introduces complex control flow more suitable for CPUs. However, many real-world sparse GPU workloads involve sparse-dense operations, where only one operand is sparse and the rest are dense. Existing sparse compilers often  fail to fully optimize the dense computation components common in these workloads, resulting in poor performance.


In this work, we build on the idea of using indirect indexing in Einsums to express sparse computation. Our contribution lies in applying it to encode format information directly into the Einsum, enabling sparse operations to be expressed in a \textit{format-conscious} manner. For example, an elementwise multiplication in COO format is expressed as the following indirect Einsum : $C_{AI_p} = AV_p \cdot B_{AI_p}$\footnote{sometimes expressed as \texttt{C[AI[p]] = AV[p] * B[AI[p]]}.}
where $AV$ contains the nonzero values and $AI$ holds their coordinates. This operation gathers elements from $B$ at positions specified by $AI$ and scatters the results back to $C$.

On top of this, we develop \textbf{Insum}, a compiler that lowers these indirect Einsums into high-performance GPU kernels by reusing existing dense tensor compilers. Unlike traditional sparse compilers, which focus on CPU codegen and sparse-specific primitives, Insum leverages dense compilation infrastructure while preserving support for irregular memory access patterns. 
By fusing indirect accesses with computation, Insum generates sparse GPU kernels from a single high-level statement, bridging the gap between sparse abstraction and dense-level performance.

This paper's contributions are as follows:
\begin{itemize}[leftmargin=10pt]
\item We show how to implement format-agnostic Einsums over sparse tensors by rewriting them into \textit{Indirect Einsums}, which map the data and metadata of sparse tensor formats onto dense tensor operations through indirect indexing.

\item We propose \textit{GroupCOO} and \textit{BlockGroupCOO}, new sparse formats enabling efficient indirect Einsums.

\item  We present a compiler for indirect Einsums, called \textit{Insum}, which lowers these expressions to dense tensor operations. In this work, we use the PyTorch~\cite{paszke2019pytorch,pytorch2.0} as the backend, targeting its FX graph~\cite{fx} intermediate representation.

\item  We extend PyTorch compiler for better generated code: \begin{itemize} 
\item  We enable pattern matching on expressions to generate direct Tensor Core operations, eliminating the need for hand-written templates.
\item  We add a codegen technique called \textit{Lazy Broadcasting} to improve performance when using Tensor Cores. 
\end{itemize}

\item Our compiler achieves competitive or superior performance across diverse sparse applications, while requiring significantly fewer lines of code (LoC), as shown in Table~\ref{tab:comparison}. 

\end{itemize}

\section{Background and Related Works}

\subsection{Einsums}

The Einstein summation~\cite{einstein_grundlage_1916} provides a compact and expressive notation for describing tensor computations, where operations are expressed as products over indexed expressions. Reductions are implicitly defined over index variables that do not appear in the output expression. In other words, if an index appears in the input tensors but not in the output, the computation reduces over that index. For example, matrix multiplication is expressed as $C_{m,n} = A_{m,k} * B_{k,n}$, where the computation reduces over the index $k$.

In recent years, Einsum notation has gained widespread popularity~\cite{harris_array_2020, ragan-kelley_halide_2013, kjolstad_tensor_2017,nayak2023teaal,odemuyiwa_edge_2024,jax2018github, paszke2019pytorch, tvm}, with generalizations that support arbitrary pointwise operations and reduction operators (e.g., \texttt{sum}, \texttt{min}, \texttt{max}) over tensor accesses with complex index expressions, including affine~\cite{tacoucf} or indirect indexing~\cite{ragan-kelley_halide_2013}. Einsums define computation as a traversal over all combinations of index variables, making it a powerful abstraction for dense tensor operations~\cite{nayak2023teaal}.

Importantly, Einsums are not tied to any particular data representation—it serves as an abstract language for expressing computation between tensors. Prior sparse frameworks~\cite{kjolstad_tensor_2017,finch,mlirsparse,nayak2023teaal} leverage this property by decoupling the Einsum specification from the storage format~\cite{chou2018format}, allowing users to select appropriate sparse representations for each tensor. In this work, we focus on \textit{indirect Einsums}, an extension that allows tensor access to appear within the index expressions of other tensors.

\subsection{Implementation of Einsums}

\subsubsection{Dense Tensor Compilers} \label{subrelated}
A number of dense tensor compilers support high-performance tensor programs using Einsum notation as input. 
Systems like Halide~\cite{ragan-kelley_halide_2013} and TVM~\cite{tvm} introduce scheduling languages that decouple the algorithm from its execution strategy. These frameworks support loop-level transformations such as \texttt{split}, \texttt{fuse}, and \texttt{reorder} to express a wide variety of implementations. 

Deep learning frameworks such as PyTorch~\cite{paszke2019pytorch}, NumPy~\cite{harris_array_2020}, and JAX~\cite{jax2018github} also provide native support for Einsum primitives. PyTorch can compile Einsum expressions through the PyTorch compiler stack~\cite{pytorch2.0}, which eventually compiles to Triton~\cite{tillet2019triton} for GPU execution. Unlike Halide or TVM, the PyTorch compiler does not expose a full scheduling language; instead, it applies a fixed set of heuristics and autotuning over a small search space (e.g., tile size). This design allows for fast compilation while still achieving competitive or better performance than other existing tensor compilers~\cite{pytorch2.0}.


\subsubsection{Sparse Tensor Compilers}
Sparse tensor compilers such as TACO~\cite{kjolstad_tensor_2017}, Finch~\cite{finch}, and \texttt{mlir-sparse}~\cite{mlirsparse} take two inputs: an Einsum and a sparse format specification for each operand. To skip ineffectual (zero) computations, these compilers must generate complex control flow for intersection and union patterns over sparse indices—capabilities not supported by most dense tensor compilers. As a result, these systems often build their own code generation frameworks from scratch, using techniques such as \textit{Looplets}~\cite{ahrens2023looplets} or \textit{merge lattices}~\cite{kjolstad_tensor_2017}. Due to the inherent control-flow complexity, most of these compilers target CPU code generation and tend to under-optimize the dense portions of computation.

Some recent efforts, such as COMET~\cite{comet} and SparseTIR~\cite{ye2023sparsetir}, aim to build sparse compilers using existing dense compiler infrastructures. COMET, built on MLIR~\cite{mlir}, primarily targets CPU code generation. SparseTIR, built on TVM~\cite{tvm}, can generate high-performance sparse GPU kernels using TVM's code generation backend and is the most closely related to our work. However, SparseTIR still requires manual scheduling, which demands significant user effort and expertise. We thoroughly compare our approach against existing GPU sparse compilers, SparseTIR and TACO, in Section~\ref{subsection:sparsecompiler}.



\section{Overview}

\begin{figure}[t]
    \centering
    \includegraphics[width=.9\linewidth]{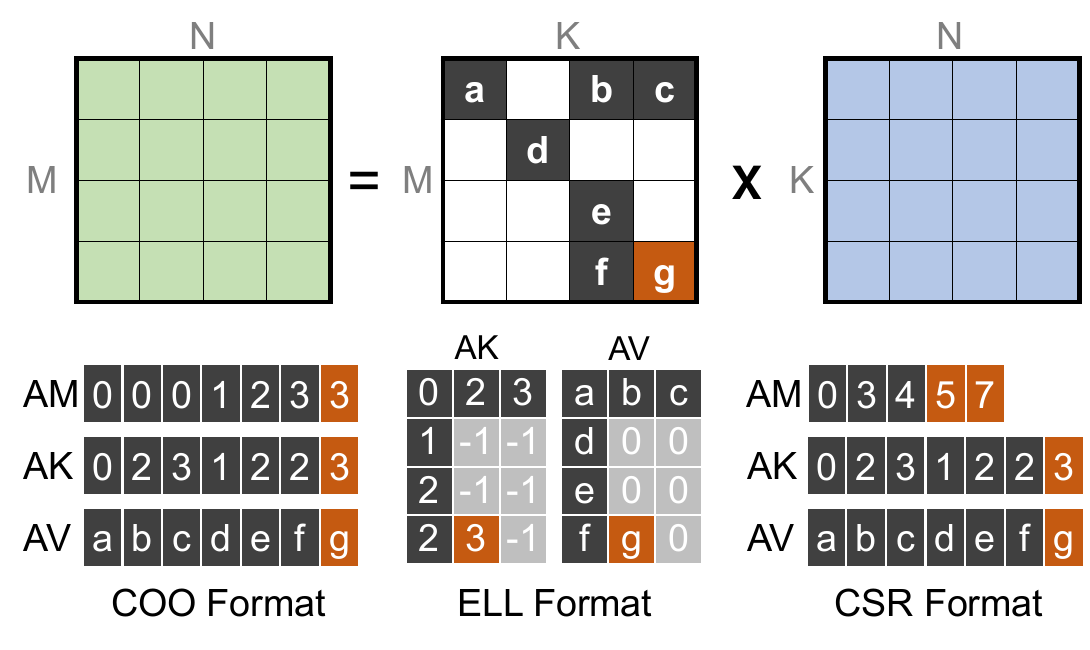}
    \caption{SpMM example ($C_{m,n} = A_{m,k} * B_{k,n}$) and various sparse formats for the matrix $A$: COO, ELL, and CSR.}
    \label{fig:spmmformat}
\end{figure}

\subsection{Extending Einsums with Indirect Indexing}

In this work, we use an extended Einsum notation that supports indirect indexing—indexing via values from other tensors. For example, we allow expressions like $C_{X_i} = A_{Y_i,j} * B_j$, where $C_{X_i}$ and $A_{Y_i,j}$ are indirect accesses. 


We can implement indirect indexing on the right-hand side using a gather, and on the left-hand side using a scatter. Similar to traditional Einsums, we assume that the output tensor is implicitly initialized to zero, and multiple writes to the same location during the scatter are resolved by summation. 
This behavior is consistent with the operational semantics of Einsum described in prior work, where overlapping contributions to the same output position are accumulated~\cite{nayak2023teaal}.



\textbf{Sparse Computation in Indirect Einsums.}
To avoid ineffectual operations ($a*0=0$), sparse tensors are often stored in compressed formats that record only the nonzero values along with metadata describing their coordinates as shown in Figure~\ref{fig:spmmformat}. These formats make it easier to identify and skip ineffectual operations.
The key idea of this paper is to express sparse computations by mapping both the data and metadata of sparse representations onto dense tensor operations using indirect indexing. In doing so, we convert a Einsum originally operating over sparse data into an indirect Einsum that operates entirely over dense tensors.

\begin{figure}[t]
    \centering
    \includegraphics[width=.8\linewidth]{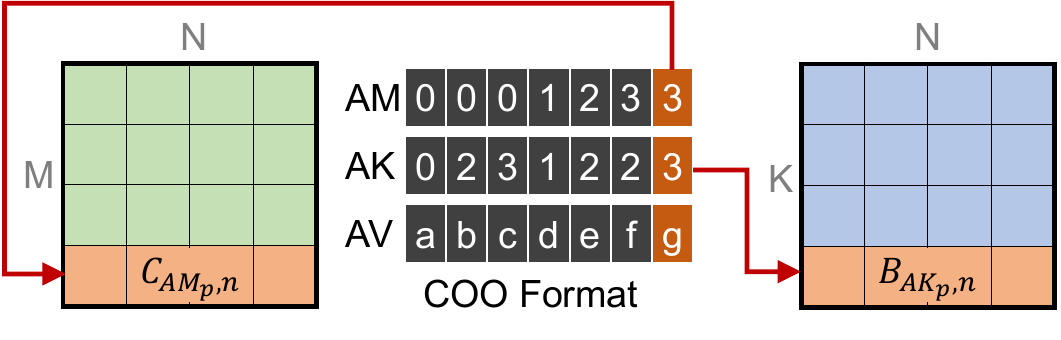}
    \caption{COO SpMM : $C_{AM_p, n} = AV_p * B_{AK_p, n}$}
    \label{fig:coospmm}
\end{figure}

Consider a matrix multiplication expressed in Einsum notation as $C_{m,n} = A_{m,k} * B_{k,n}$, where $A$ is sparse and $B$ is dense, creating a sparse-dense matrix multiply (SpMM). To avoid storing and multiplying by zeros in $A$, one can leverage a sparse format such as COO, which stores only the nonzero values of $A$ in a vector \texttt{AV}, along with their corresponding row and column coordinates in \texttt{AM} and \texttt{AK}, respectively. Using this representation, we access only the elements of $B$ that correspond to the nonzero coordinates in $A$, avoiding ineffectual computation. This strategy corresponds to a leader-follower intersection as described in prior works~\cite{ahrens2023looplets,sparseloop}. This operation can be expressed using an indirect Einsum by accumulating into $C$ as Figure~\ref{fig:coospmm}.

\begin{figure*}[t]
  \centering
  \includegraphics[width=\linewidth]{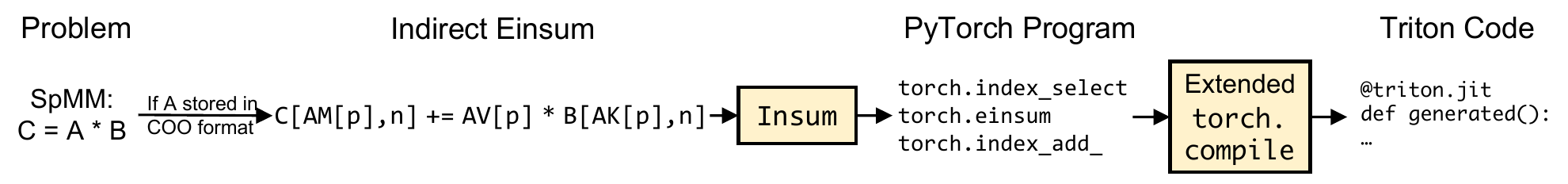}
  \caption{Overview of our system. Given a sparse problem, users express it in an indirect Einsum. This expression is then translated into PyTorch program using \texttt{Insum} compiler, and then lowered into a Triton kernel via our extended \texttt{torch.compile}.}
  \label{fig:overview}
\end{figure*}

\subsection{Compiler Overview}


Figure~\ref{fig:overview} illustrates the workflow of our compiler. First, users express sparse computations using indirect Einsums (Section~\ref{sec:format}). 
To execute indirect Einsums on a GPU, we developed the \texttt{Insum} compiler, which leverages the existing compiler for dense tensor operations.
\texttt{Insum} lowers the indirect Einsum expression into a semantically equivalent PyTorch program using PyTorch primitives such as \texttt{torch.index\_select} and \texttt{torch.einsum} (Section~\ref{subsec:insum}). Finally, this program is passed to our extended PyTorch compiler (Section~\ref{subsec:extended}).




\section{Fixed-Length Sparse Formats} \label{sec:format}

While the previous section explains how to express sparse computations in COO format using indirect Einsums, not all sparse formats are directly compatible with this approach. A fundamental limitation of the Einsum-based expression model is that the loop bounds for each index variable must be fixed and cannot depend on data values. Many sparse formats, such as CSR format, require data-dependent loop bounds. For instance, SpMM CSR in Figure~\ref{fig:spmmformat} iterates over rows, and within each row, it iterates over a variable number of nonzeros encoded in \texttt{AM}, as shown below:

\begin{lstlisting}[basicstyle=\ttfamily\footnotesize]
for m in 0..M: 
  for p in AM[m]..AM[m+1]: # variable-length loop
    for n in 0..N: 
      C[m,n] += Av[p] * B[AK[p],n]
\end{lstlisting}

Since this kind of variable-length loop cannot be directly expressed in Einsums, we must either store the data in a format like COO or pad nonzeros per row to have uniform length like ELL~\cite{ell}, which is compatible with fixed loop bounds. We refer to such formats as \emph{fixed-length formats}.

The ELL format has the advantage of avoiding explicit storage of coordinates along the row dimension, thereby eliminating the need for scatter operations. However, it can introduce significant padding overhead when the nonzero distribution is uneven. To address this, we propose a fixed-length sparse format called \textbf{GroupCOO}, which strikes a balance between COO and ELL by grouping nonzeros in the same dimension. This results in improved performance by reducing padding compared to ELL and requiring less scatter than COO. Additionally, we demonstrate that GroupCOO can be extended to support block-sparse formats as well. 


\subsection{Grouping} 

We introduce \textbf{grouping}, a technique that helps derive various GroupCOO formats. To illustrate this, we walk through an example of SpMM, $C_{m,n} = A_{m,k} * B_{k,n}$, where \texttt{A} is sparse.

\begin{figure}[h]
    \centering
    \includegraphics[width=\linewidth]{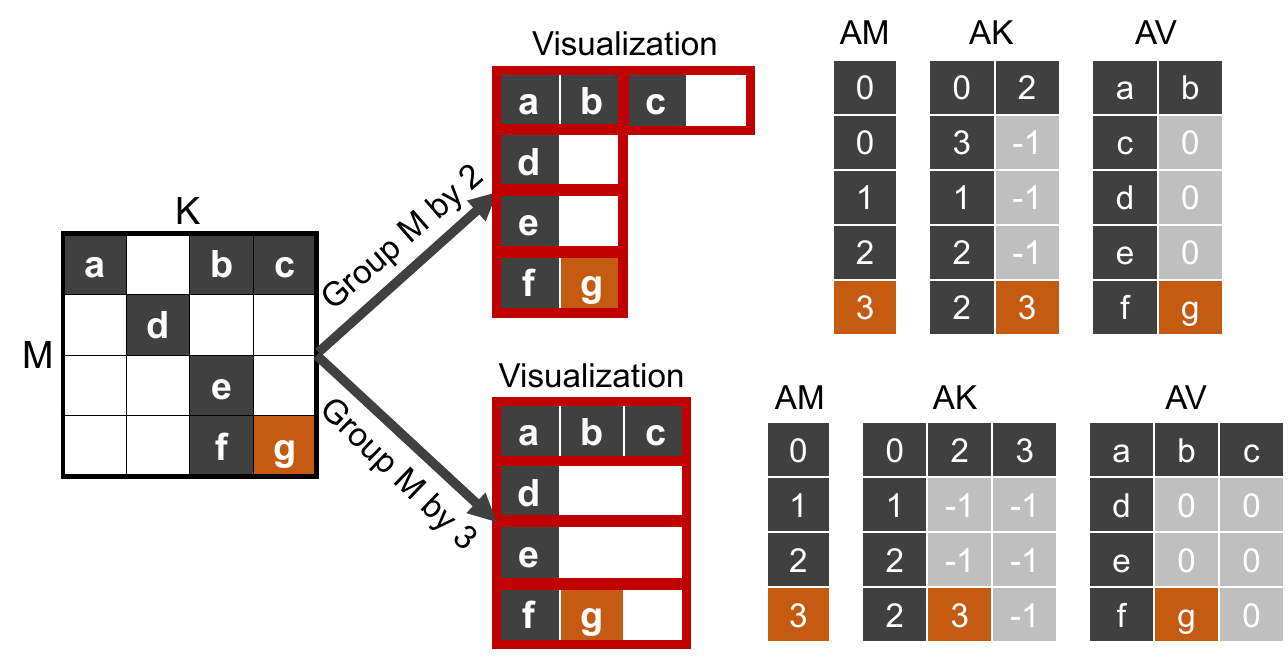}
    \caption{GroupCOO format with grouping on the \texttt{M} dimension by group sizes 2 and 3. Grouping with the maximum number of nonzero per row (size = 3) yields the ELL format.}
    \label{fig:groupcoo}
\end{figure}

\textbf{Grouping:} While the COO format is simple, it can be inefficient when the same row or column indices appear repeatedly. To reduce this redundancy, we introduce \textbf{GroupCOO}, a fixed-length format that partitions nonzeros into \textbf{groups} along a chosen dimension (e.g., rows) and stores the group index only once per group. For example, in Figure~\ref{fig:groupcoo}, we group the nonzeros along the \texttt{M} dimension (rows) using group sizes of 2 and 3. Within each group, we pad the data if necessary and this padding makes the format fixed-length while reducing redundancy of the same coordinates.

The GroupCOO SpMM can be written as follows, where \texttt{p} iterates over groups and \texttt{q} over elements within each group:
\begin{equation*}
C_{AM_p,\,n} = AV_{p,q} * B_{AK_{p,q},\,n}
\end{equation*}

GroupCOO sits between the COO format and the ELL format: setting group size to 1 degenerates to the original COO format, while setting group size to the maximum number of nonzeros per row yields the ELL format (Figure~\ref{fig:groupcoo}). 

\begin{figure}[h]
    \centering
    \includegraphics[width=.8\linewidth]{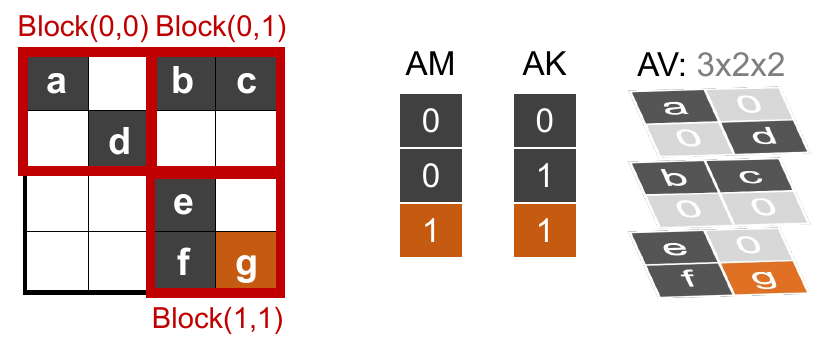}
    \caption{BlockCOO SpMM: $C_{AM_p,\,bm,\,n} = AV_{p,\,bm,\,bk} * B_{AK_p,\,bk,\,n}$, where \texttt{bm} and \texttt{bk} index within each block.}
    \label{fig:blockcoo}
\end{figure}

\textbf{Group+Blocking:}
Blocking is a common technique in sparse computation to exploit local dense patterns by dividing a sparse matrix into dense blocks~\cite{chou2018format,won2023waco,quansight2023bsr}. The BlockCOO format stores the coordinates of each nonzero block in \texttt{AM} (row block) and \texttt{AK} (column block), and the block values in \texttt{AV} as a dense tensor of shape \texttt{[num\_blocks, bM, bK]}. SpMM in BlockCOO is expressed as Figure~\ref{fig:blockcoo}.

\begin{figure}[h]
    \centering
    \includegraphics[width=\linewidth]{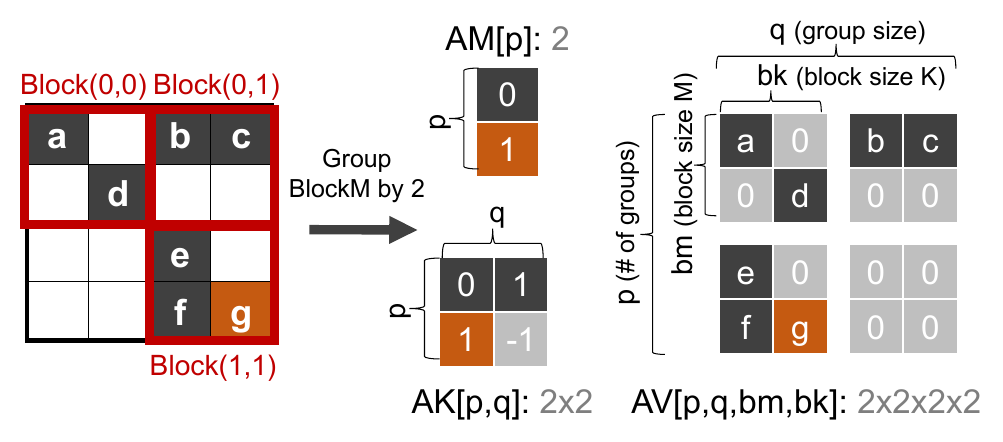}
    \caption{BlockGroupCOO SpMM : $C_{AM_p,\,bm,\,n} = AV_{p,\,q,\,bm,\,bk} * B_{AK_{p,\,q},\,bk,\,n}$, where \texttt{p} indexes groups, \texttt{q} entries within each group, and \texttt{bm}, \texttt{bk} index each block.}
    \label{fig:blockgroupcoo}
\end{figure}

Grouping can also be applied to block formats. In \textbf{BlockGroupCOO}, we group along one block-coordinate dimension, so that \texttt{AV} has shape \texttt{[num\_groups, group\_size, bM, bK]}, and \texttt{AM}/\texttt{AK} store group and block-level coordinates. SpMM in BlockGroupCOO is expressed as Figure~\ref{fig:blockgroupcoo}.

\subsection{Choosing Group Size} \label{subsec:groupsize}

\begin{figure}[t]
    \centering
    \begin{subfigure}{0.495\linewidth}
        \centering
        \includegraphics[width=\linewidth]{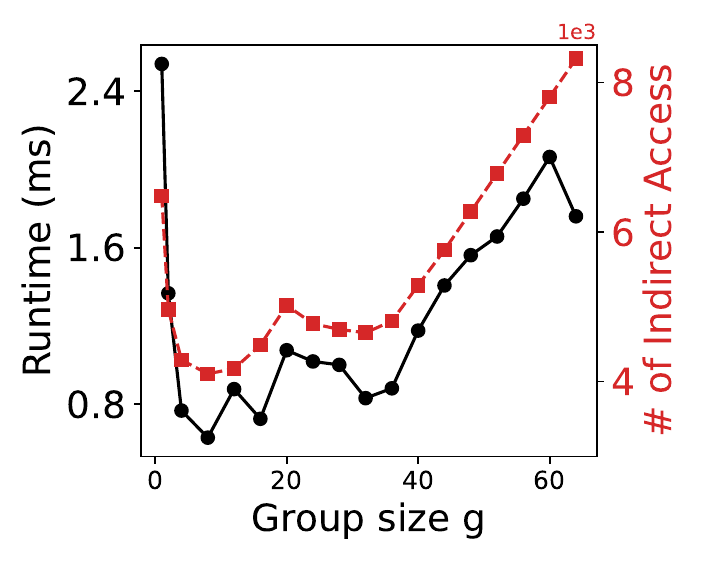}
        \caption{Runtime vs. \# of Ind. Access}
        \label{fig:ik}
    \end{subfigure}
    \begin{subfigure}{0.495\linewidth}
        \centering
        \includegraphics[width=\linewidth]{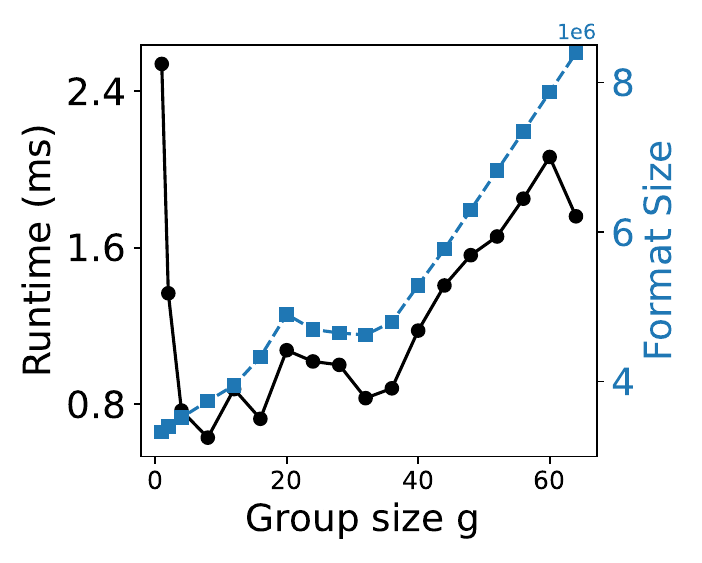}
        \caption{Runtime vs. Format Size}
        \label{fig:ikv}
    \end{subfigure}
    \caption{Runtime correlation with indirect access and size}
    \label{fig:ik_comparison}
\end{figure}

Choosing an appropriate group size~$g$ is crucial for both storage footprint and compute efficiency. If $g$ is too large, each row is padded to the next multiple of~$g$, leading to wasted memory. If $g$ is too small, the format degenerates into standard COO, losing the benefits of grouping. Ideally, we want to choose the group size~$g$ that yields the best runtime.

Our experiments suggest that minimizing \emph{indirect memory accesses}, i.e., gathers and scatters, is the key to achieving high performance. These indirect accesses often result in expensive DRAM transactions due to poor spatial locality. In Figure~\ref{fig:ik_comparison}, we run BlockGroupCOO SpMM on a $4096 \times 4096$ block-sparse matrix with $32 \times 32$ dense blocks at 80\% sparsity, sweeping various group sizes~$g$.

Another proxy we tested for the optimal~$g$ is the one that minimizes total storage (i.e., the size of AM, AK, and AV in Figure~\ref{fig:blockgroupcoo}), thereby reducing memory footprint. However, as shown in Figure~\ref{fig:ikv}, the format memory increases almost monotonically with larger $g$, especially in BlockGroupCOO, where padding in AV is influenced by block size. So, we found runtime does not correlate well with the format size.

In contrast, Figure~\ref{fig:ik} demonstrates that runtime is closely aligned with the total number of \emph{gathers and scatters} (i.e., accesses to AM and AK), which  defined as $F(g)$:
\[
F(g) =
\underbrace{
\sum_{i=0}^{n-1}
\left\lceil \frac{\mathit{occ}_i}{g} \right\rceil
}_{\text{AM: Scatter}}
+
\underbrace{
g \sum_{i=0}^{n-1}
\left\lceil \frac{\mathit{occ}_i}{g} \right\rceil
}_{\text{AK: Gather}}
=
(g + 1)
\sum_{i=0}^{n-1}
\left\lceil \frac{\mathit{occ}_i}{g} \right\rceil.
\]

Here, $n$ is the number of rows in the matrix and $\mathit{occ}_i$ is the number of nonzeros in row~$i$ (Figure~\ref{fig:groupcoo} gives $\mathit{occ} = [3,1,1,2]$). With group size~$g$, each row $i$ is divided into $\lceil \mathit{occ}_i/g\rceil$ groups.

Finding the exact optimal $g$ over the range $[1, \max_i \mathit{occ}_i]$ requires brute-force evaluation of the $F(g)$ above, leading to $O(n \cdot \max_i \mathit{occ}_i)$ complexity. To obtain a much faster, yet nearly optimal estimate, we replace the ceiling with a conservative approximation: $\left\lceil \mathit{occ}_i / g \right\rceil \approx \mathit{occ}_i / g + 1$.

Letting $S = \sum_i \mathit{occ}_i$, the relaxed cost becomes:
\[
\tilde F(g) = (g + 1)\left( \frac{S}{g} + n \right) = S + \frac{S}{g} + ng + n.
\]
Treating $g$ as a real variable and differentiating:
\[
\frac{d\tilde F}{dg} = -\frac{S}{g^2} + n = 0
\quad \Rightarrow \quad
g^\star = \sqrt{\frac{S}{n}}.
\]

This estimate provides a good approximation to the group size that minimizes indirect accesses and likely achieves optimal runtime. In practice, we round $g^\star$ to the nearest power-of-two values  and select the one with the best runtime. This choice is motivated by the fact that Triton, our backend, performs best with power-of-two block sizes, as evidenced by the downward spikes in Figure~\ref{fig:ik_comparison}. This heuristic consistently yields efficient configurations across applications (Section~\ref{section:evaluation}).

\section{Compiler Implementation} \label{sec:compiler}

This section describes our approach for generating efficient GPU code from indirect Einsum expressions. We leverage a dense tensor compiler that produces high-performance code and supports gather/scatter operations. Among available options, we selected the PyTorch compiler~\cite{pytorch2.0} for four reasons: (1) it is Python-native and user-friendly; (2) it supports automatic differentiation, essential for sparse deep learning; (3) it targets Triton, enabling Tensor Core support without manual scheduling~\cite{tillet2019triton}; and (4) it delivers state-of-the-art performance among existing dense tensor compilers~\cite{pytorch2.0}.

\subsection{Insum: Rewriting Indirect Einsums in PyTorch} \label{subsec:insum}

To use the PyTorch compiler effectively, we first need to translate indirect Einsum expressions into an equivalent PyTorch program. To facilitate this, we introduce the \texttt{Insum} compiler, which converts indirect Einsum strings into standard PyTorch operations. The interface for \texttt{Insum} is as follows:

\begin{center}
\noindent\texttt{Insum(expression: str, **tensors: dict)}
\end{center}

Here, \texttt{expression} is a string representing the indirect Einsum computation, and \texttt{tensors} is a dictionary mapping tensor names to their corresponding PyTorch tensors. For example, an indirect Einsum $C_{D_y,x} = A_{y,E_r} * B_{r,x}$ can be expressed as \texttt{Insum("C[D[y],x] += A[y,E[r]] * B[r,x]", C = C, A = A, B = B, D = D, E = E)}. 


The \texttt{Insum} function works by parsing the provided string expression and constructing an intermediate representation known as an FX graph~\cite{fx}. In the PyTorch compiler stack, an FX graph is a functional representation of a PyTorch program that clearly encodes tensor operations and their dependencies. Since PyTorch lacks a native primitive for indirect Einsums, we need to translate it into equivalent PyTorch operations. The basic workflow for this translation involves three main steps:

\begin{enumerate}[leftmargin=10pt]
    \item \textbf{Gather Inputs}: If the right-hand side involves indirect indexing, gather elements from required tensors using operations like \texttt{torch.index\_select} or \texttt{torch.gather}.
    
    \item \textbf{Execute Einsum}: Perform the Einsum (i.e., \texttt{torch.einsum}) computation using the gathered data.
    
    \item \textbf{Scatter Outputs}: If the left-hand-side expression involves indirect indexing, scatter the computed results to their target locations using operations like \texttt{torch.index\_add}.
\end{enumerate}

For the above example, \texttt{Insum} will create an FX graph equivalent to the following PyTorch program:

\begin{center}
\begin{minipage}{0.9\linewidth}
\centering
\begin{lstlisting}[basicstyle=\ttfamily\footnotesize]
# (1) Gather: Atmp[y,r] = A[y,E[r]]
Atmp = torch.index_select(A, dim=1, index=E)
# (2) Einsum: Ctmp[y,x] = Atmp[y,r] * B[r,x]
Ctmp = torch.einsum("yr,rx->yx", Atmp, B)
# (3) Scatter: C[D[y],x] += Ctmp[y,x]
C.index_add_(dim=0, index=D, source=Ctmp)
\end{lstlisting}
\end{minipage}
\end{center}

\subsection{Native Tensor Core Support in TorchInductor}\label{subsec:extended}
Once we construct an FX graph, we pass it to TorchInductor—the code generation backend of the PyTorch compiler. TorchInductor lowers the FX graph into optimized code by first translating it into an intermediate representation called \textit{InductorIR}, a loop-level IR that describes how operations are executed through nested loops. At this level, TorchInductor performs optimizations such as loop tiling, reordering, and fusion before generating final Triton codes.

\textbf{Limitation:} A core optimization in TorchInductor is the \textit{fusion} of multiple PyTorch operations into a single loop nest. TorchInductor can fuse many pointwise and reduction operations into a single Triton kernel. However, there is a significant exception: matrix multiplication.

Due to its critical role in deep learning workloads, matrix multiplication is not generated natively by TorchInductor. Instead, it relies on a hand-optimized Triton template. While this template allows limited fusion of pointwise operations, it prevents complex operations, such as gather and scatter, from being fused. This lack of fusion can result in increased memory traffic, reduced locality, and degraded performance for computations involving matrix multiplication.

When the FX graph for \texttt{"C[D[y],x] += A[y,E[r]] * B[r,x]"} is passed to TorchInductor, it generates three separate (i.e., not fused) Triton kernels: one for gathering \texttt{A}, one using a matrix multiplication template (\texttt{Ctmp[y,x] = Atmp[y,r] * B[r,x]}), and one for scattering to \texttt{C}.

To overcome the limitations of fixed templates, we extend TorchInductor’s codegen pass to emit the Triton \texttt{tl.dot} intrinsic, which maps directly to Tensor Core operations. This enables the compiler to natively generate Tensor Core optimized kernels without relying on hand-written templates. Implemented directly on TorchInductor’s loop-level IR, our approach integrates seamlessly with its fusion infrastructure, allowing fully fused kernels for indirect Einsums. 


\begin{figure*}[htbp]
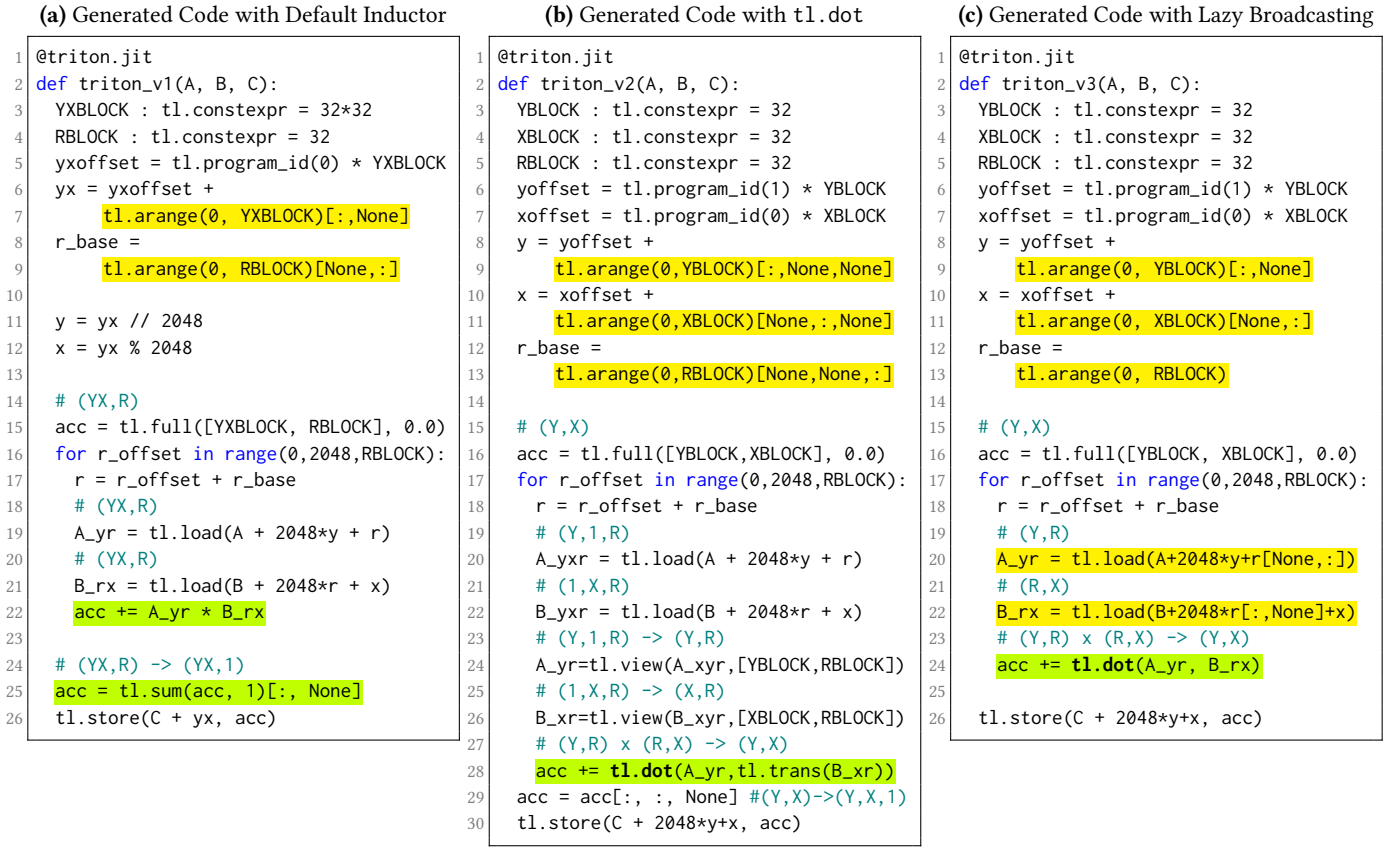

    \centering
    \begin{subfigure}[t]{0.31\textwidth}
        \centering
        \caption{Generated Code with Default Inductor}
        \begin{lstlisting}[basicstyle=\ttfamily\footnotesize]
@triton.jit
def triton_v1(A, B, C):
  YXBLOCK : tl.constexpr = 32*32
  RBLOCK : tl.constexpr = 32
  yxoffset = tl.program_id(0) * YXBLOCK
  yx = yxoffset +
       (*@\hl{tl.arange(0, YXBLOCK)[:,None]} @*)
  r_base =
       (*@\hl{tl.arange(0, RBLOCK)[None,:]} @*)

  y = yx // 2048
  x = yx % 2048

  # (YX,R)
  acc = tl.full([YXBLOCK, RBLOCK], 0.0)
  for r_offset in range(0,2048,RBLOCK):
    r = r_offset + r_base
    # (YX,R)
    A_yr = tl.load(A + 2048*y + r) 
    # (YX,R)
    B_rx = tl.load(B + 2048*r + x) 
    (*@\hlgreen{acc += A\_yr * B\_rx}@*)
    
  # (YX,R) -> (YX,1)
  (*@\hlgreen{acc = tl.sum(acc, 1)[:, None]}@*)
  tl.store(C + yx, acc)
        \end{lstlisting}
        \label{fig:codev1}
    \end{subfigure}
    \hfill
    \begin{subfigure}[t]{0.31\textwidth}
        \centering
        \caption{Generated Code with \texttt{tl.dot}}
        \begin{lstlisting}[basicstyle=\ttfamily\footnotesize]
@triton.jit
def triton_v2(A, B, C):
  YBLOCK : tl.constexpr = 32
  XBLOCK : tl.constexpr = 32
  RBLOCK : tl.constexpr = 32
  yoffset = tl.program_id(1) * YBLOCK
  xoffset = tl.program_id(0) * XBLOCK
  y = yoffset + 
      (*@\hl{tl.arange(0,YBLOCK)[:,None,None]} @*)
  x = xoffset +
      (*@\hl{tl.arange(0,XBLOCK)[None,:,None]} @*)
  r_base =
      (*@\hl{tl.arange(0,RBLOCK)[None,None,:]} @*)

  # (Y,X)
  acc = tl.full([YBLOCK,XBLOCK], 0.0)
  for r_offset in range(0,2048,RBLOCK):
    r = r_offset + r_base
    # (Y,1,R)
    A_yxr = tl.load(A + 2048*y + r) 
    # (1,X,R)
    B_yxr = tl.load(B + 2048*r + x) 
    # (Y,1,R) -> (Y,R)
    A_yr=tl.view(A_xyr,[YBLOCK,RBLOCK])
    # (1,X,R) -> (X,R)
    B_xr=tl.view(B_xyr,[XBLOCK,RBLOCK])
    # (Y,R) x (R,X) -> (Y,X)
    (*@\hlgreen{acc += \textbf{tl.dot}(A\_yr,tl.trans(B\_xr))}@*)
  acc = acc[:, :, None]  #(Y,X)->(Y,X,1)  
  tl.store(C + 2048*y+x, acc)
        \end{lstlisting}
        \label{fig:codev2}
    \end{subfigure}
    \hfill
    \begin{subfigure}[t]{0.31\textwidth}
        \centering
        \caption{Generated Code with Lazy Broadcasting}
        \begin{lstlisting}[basicstyle=\ttfamily\footnotesize]
@triton.jit
def triton_v3(A, B, C):
  YBLOCK : tl.constexpr = 32
  XBLOCK : tl.constexpr = 32
  RBLOCK : tl.constexpr = 32
  yoffset = tl.program_id(1) * YBLOCK
  xoffset = tl.program_id(0) * XBLOCK
  y = yoffset +
      (*@\hl{tl.arange(0, YBLOCK)[:,None]} @*) 
  x = xoffset +
      (*@\hl{tl.arange(0, XBLOCK)[None,:]} @*) 
  r_base =
      (*@\hl{tl.arange(0, RBLOCK)} @*) 
      
  # (Y,X)
  acc = tl.full([YBLOCK, XBLOCK], 0.0)
  for r_offset in range(0,2048,RBLOCK):
    r = r_offset + r_base
    # (Y,R)
    (*@\hl{A\_yr = tl.load(A+2048*y+r[None,:])} @*) 
    # (R,X)
    (*@\hl{B\_rx = tl.load(B+2048*r[:,None]+x)} @*) 
    # (Y,R) x (R,X) -> (Y,X)
    (*@\hlgreen{acc += \textbf{tl.dot}(A\_yr, B\_rx)}@*)

  tl.store(C + 2048*y+x, acc)
        \end{lstlisting}
        \label{fig:codev3}
    \end{subfigure}
\caption{
Triton code generated by TorchInductor for the matrix multiply between two 2048 $\times$ 2048 matrices, \texttt{C[y,x] = A[y,r] * B[r,x]}. 
(a) Default code generation without our compiler optimization. (b) Code generation with \texttt{tl.dot}: introduces a new \texttt{ops.dot} IR node that enables Tensor Core and applies 2D tiling over output dimensions. 
(c) Code generation with Lazy Broadcasting: eliminates redundant reshaping and broadcasting by delaying layout expansion until needed.}
    \label{fig:triton-kernels}
\end{figure*}

\subsubsection{Forcing TorchInductor to Generate Matmul}

While TorchInductor defaults to a hand-written template for matrix multiplication, it can also be configured to generate it natively. We build on this capability in our compiler extension.

Consider the PyTorch matrix multiplication example (\texttt{C = torch.matmul(A,B)}) with matrices of shape $(2048, 2048)$. To prompt TorchInductor to generate matrix multiplication kernels without a template, a useful trick is to rewrite the operation explicitly as replicated multiplication (i.e., batched Cartesian product) followed by summation :


\begin{lstlisting}[basicstyle=\ttfamily\footnotesize]
# Prod[y,r,x] = A[y,r] * B[r,x]
Prod = A.unsqueeze(2) * B.unsqueeze(0)
# C[y,x] = Prod[y,r,x]
C = Prod.sum(dim=1)  
\end{lstlisting}

When provided with this rewritten computation, TorchInductor generates the Triton kernel depicted in Figure~\ref{fig:codev1}. However, the resulting kernel suffers from performance limitations due to two critical issues:

\begin{itemize}[leftmargin=12pt]
\item {\bf Lack of Tensor Core Utilization:}
TorchInductor generates naive multiplication and summation (as seen on the green-highlighted code) instead of leveraging Triton's optimized {\tt tl.dot} operation for tensor cores.

\item {\bf Absence of Proper Tiling:}
TorchInductor currently flattens all pointwise indices—those not involved in the reduction—into a single loop dimension. This approach often benefits from the GPU's abundant parallelism by enlarging the parallelizable loop. However, it can also negatively impact memory locality. In particular, output dimensions $y$ and $x$, which are critical for memory access patterns in matrix multiplication, are flattened together. This behavior is evident in the loop-level InductorIR:

\begin{lstlisting}[basicstyle=\ttfamily\footnotesize]
for yx:
  y, x = yx // 2048, yx % 2048
  for r:
    Prod[y,r,x] = A[y,r] * B[r,x] # multiply
    C[y,x] = Prod[y,r,x]          # summation
\end{lstlisting}


\end{itemize}

\subsubsection{Invoking Tensor Cores with \texttt{tl.dot}}


To achieve optimal performance on GPUs, the generated code must leverage Triton’s intrinsic \texttt{tl.dot} and apply efficient two-dimensional tiling over the output matrix. To support this, we introduce a new IR node in InductorIR called \texttt{ops.dot}, which explicitly represents matrix multiplication. 

Although \texttt{ops.dot} has the same semantics as broadcasted multiplication followed by summation, it is lowered differently during code generation: it maps directly to \texttt{tl.dot}. We implemented a rewrite pass that detects patterns of broadcasted multiplication followed by summation, and replaces them with an \texttt{ops.dot} node.

When this node is present, Inductor explicitly tiles the computation along the pointwise variables \( y \) and \( x \), producing the following tiled IR:

\begin{lstlisting}[basicstyle=\ttfamily\footnotesize]
for y:
  for x:
    for r:
      # Equivalent to C[y,x] += A[y,r] * B[r,x]
      ops.dot(C[y, x], A[y, r], B[r, x])  
\end{lstlisting}


Figure~\ref{fig:codev2} shows the generated code using the tiled IR with \texttt{ops.dot}. Compared to the naive version in Figure~\ref{fig:codev1}, there are two notable differences:

\begin{enumerate}[leftmargin=12pt]
    \item \textbf{Tiled $y$ and $x$ (Lines 8--13):} The computation is now tiled over three explicit axes: \( y \), \( x \), and \( r \), with \( y \) and \( x \) parallelized. This replaces the earlier flattened \texttt{yx} index.

    \item \textbf{Tensor Core Invocation via \texttt{tl.dot} (Lines 23--28):} Instead of manually multiplying and summing over the reduction axis, Inductor now emits a \texttt{tl.dot} operation from \texttt{ops.dot}, enabling the use of Tensor Cores. The \texttt{tl.dot} intrinsic expects its operands to have shapes \((Y, R)\) and \((R, X)\), producing an output of shape \((Y, X)\). To meet these shape requirements, the input tensors are first reshaped to \((\texttt{YBLOCK}, \texttt{RBLOCK})\) and \((\texttt{RBLOCK}, \texttt{XBLOCK})\) using \texttt{tl.view}, and the second operand is transposed with \texttt{tl.trans}. The accumulator, initialized in line 16, excludes the reduction dim \texttt{RBLOCK}, following an output-stationary dataflow to retain the partial sums directly in the output tile.
\end{enumerate}

\subsubsection{Lazy Broadcasting} \label{subsublazy}

While adding \texttt{ops.dot} and tiling the pointwise dimensions already allows us to generate a fully fused gather-scatter kernels with Tensor Core, we found that an additional optimization---\textit{Lazy Broadcasting}---is necessary to reach peak performance.

By default, when Inductor lowers IR to Triton code, it uses what we refer to as \textit{Eager Broadcasting}: every loop variable is assigned a unique axis in the Triton block dimensions up front. For example, in the yellow-highlighted code in Figure~\ref{fig:codev2}, \texttt{YBLOCK}, \texttt{XBLOCK}, and \texttt{RBLOCK} correspond to \( y \), \( x \), and \( r \), shaped as \texttt{[:,None,None]}, \texttt{[None,:,None]}, and \texttt{[None,None,:]}, respectively. This eager broadcasting is convenient for code generation, as it simplifies indexing logic---each tensor can be accessed easily without needing to reason about broadcasting behavior at load time.

However, a major downside is that \texttt{tl.dot} expects operands in a very specific layout: \((Y, R) \times (R, X)\). The eager broadcasting breaks this layout, requiring explicit reshaping and transposing before the \texttt{tl.dot} call. We found that this introduces runtime overhead on reshaping and transposing.

To eliminate this overhead, we introduce \textit{Lazy Broadcasting}. Rather than broadcasting every loop variable with a unique axis from the start, we delay broadcasting until it is actually required. This allows us to match the expected layout of \texttt{tl.dot}, without any reshaping or transposing.

Figure~\ref{fig:codev3} shows the generated code after applying Lazy Broadcasting. Two key differences from Figure~\ref{fig:codev2}:

\begin{enumerate}[leftmargin=12pt]
\item \textbf{Minimal Eager Broadcasting (Lines 8--13):} Loop variables whose broadcasting dimensions stay constant are expanded once at the beginning. Only $y$ and $x$ are shaped as \texttt{tl.arange(0, YBLOCK)[:, None]} and \texttt{tl.arange(0, XBLOCK)[None, :]}, while $r$ remains 1D.

    \item \textbf{Lazy Broadcasting (Lines 19--22):} We apply broadcasting to \( r \) on demand, depending on its usage. For example, when loading \texttt{A[y, r]}, we use \texttt{r[None, :]} to form shape \texttt{(YBLOCK,RBLOCK)}; when loading \texttt{B[r, x]}, we use \texttt{r[:, None]} to form \texttt{(RBLOCK,XBLOCK)}. This ensures that the operands passed to \texttt{tl.dot} are already in the correct shape, removing the need for views or transposes.
\end{enumerate}


To implement lazy broadcasting, we maintain the block shape metadata for every Triton variable during codegen. At kernel entry, only the output indices (\texttt{y} and \texttt{x}) are eagerly materialized as a 2D block, while the reduction index \texttt{r\_base} is kept as a 1D block. As new variables are generated, their block shape is inferred from the operands they depend on. Comments in Figure~\ref{fig:codeexample} show the block shape of each variable.

While tracking block shapes during code generation, whenever a variable with 1D shape \texttt{(RBLOCK,)} interacts with a 2D variable shaped with \texttt{YBLOCK} or \texttt{XBLOCK}, we apply lazy broadcasting to the 1D variable along the required axis. For instance, when compiling $C_{D_y,x} = A_{y,E_r} * B_{r,x}$ in Figure~\ref{fig:codeexample}, we defer broadcasting of 1D variables \texttt{r} and \texttt{E\_r} until they are needed in loading \texttt{A[y,E[r]]} or \texttt{B[r,x]}. As a result, the inputs to \texttt{tl.dot} naturally take the shape \texttt{(YBLOCK,RBLOCK)} and \texttt{(RBLOCK,XBLOCK)}, enabling direct use of Tensor Cores without requiring extra reshaping or transposition.

\begin{figure}[t]
\begin{lstlisting}[basicstyle=\ttfamily\footnotesize]
@triton.jit
def generated_triton(A, B, C):
 yoffset = tl.program_id(1) * YBLOCK
 xoffset = tl.program_id(0) * XBLOCK
    
 y = yoffset + tl.arange(0, YBLOCK)[:, None] # (Y,1)
 x = xoffset + tl.arange(0, XBLOCK)[None, :] # (1,X)
 r_base = tl.arange(0, RBLOCK) # (R,)
    
 acc = tl.full([YBLOCK, XBLOCK], 0.0) # (Y,X)
 for r_offset in range(0, 2048, RBLOCK):
  r = r_offset + r_base # (R,)
  E_r = tl.load(E + r)  # (R,)
  (*@\hl{A\_yr = tl.load(A + (E\_r[None, :] + 2048 * y))}@*) # (Y,R)
  (*@\hl{B\_rx = tl.load(B + (x + 2048 * r[:, None]))}@*) # (R,X)
  acc += tl.dot(A_yr, B_rx) # (Y,X)

 D_y = tl.load(D + y) # (Y,1)
 tl.atomic_add(C + (x + 2048 * D_y), acc) #(Y,X)
\end{lstlisting}
\caption{Generated Triton code for \texttt{C[D[y],x] += A[y,E[r]] * B[r,x]}, showing full fusion of gather, matmul, and scatter using Tensor Cores. Yellow-highlighted lines indicate where lazy broadcasting occurs.}
\label{fig:codeexample}
\end{figure}


\section{Evaluation} \label{section:evaluation}
In this section, we evaluate our generated sparse GPU kernels using \texttt{Insum} using our PyTorch compiler extension. We focus on the following five research questions:

\begin{itemize}[leftmargin=10pt]
\item \textbf{Q1:} How universal is our indirect Einsum model across different sparse applications?
\item \textbf{Q2:} How does the performance of our generated code compare to hand-written, state-of-the-art kernels?
\item \textbf{Q3:} How does grouping or blocking in the fixed-length sparse format affect performance? 
\item \textbf{Q4:} How does our new codegen affect performance?
\item \textbf{Q5:} How does our compiler compare to existing sparse GPU compilers?
\end{itemize}

We evaluate \textbf{Q1} and \textbf{Q2} by comparing our generated code against hand-optimized kernels across four domains. We address \textbf{Q3} and \textbf{Q4} through an ablation study in Section~\ref{subsection:analysis}. Finally, we answer \textbf{Q5} by comparing our compiler to existing compilers across various dimensions in Section~\ref{subsection:sparsecompiler}.


\subsection{Experimental Setup}

The \texttt{Insum} comprises approximately 500 lines of code, and modifications to TorchInductor add about 800 lines. Our implementation is based on a patched version of the PyTorch (commit \texttt{e8304f0}). All experiments are conducted on an RTX3090 (24GB, Ampere). All group sizes of the format were selected using the heuristic described in Section~\ref{subsec:groupsize}.

\textbf{Case Studies:} We evaluate our compiler across four different sparse kernels commonly used in deep learning. For each application, we compare our generated code against state-of-the-art implementations, which often involve significant engineering effort and expert-written Triton or CUDA.


\begin{itemize}[leftmargin=20pt]
    \item \textbf{Structured SpMM:} A sparse matrix-dense matrix multiplication (SpMM), where the sparse matrix exhibits structured sparsity—specifically, block sparsity. 

    \item \textbf{Unstructured SpMM:} Similar to Structured SpMM, but the sparse matrix features unstructured sparsity. 

    \item \textbf{Point Cloud Sparse Convolution:} A sparse convolution operation over 3D point clouds or sparse voxels. 

    \item \textbf{Equivariant Tensor Product:} A fully connected tensor product used in equivariant neural networks. 

\end{itemize}

\begin{figure}[t]
  \centering
    \centering
    \includegraphics[width=0.9\linewidth]{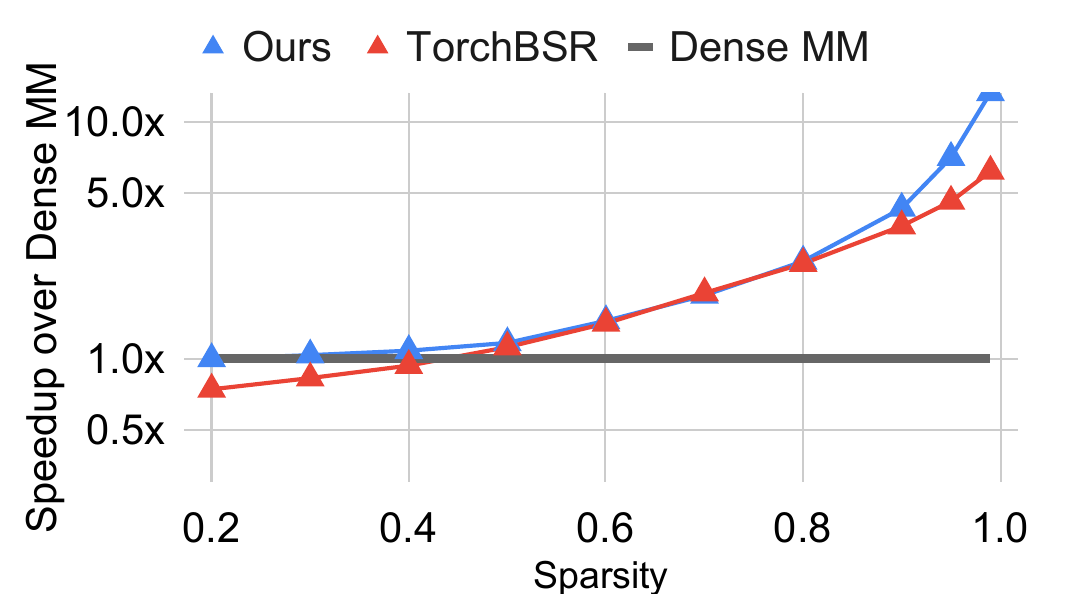}
    \caption{Speedup of our kernel vs. TorchBSR on FP16.}
  \label{fig:bspmm}
\end{figure}
\begin{figure*}[t]
  \centering
  \includegraphics[width=\linewidth]{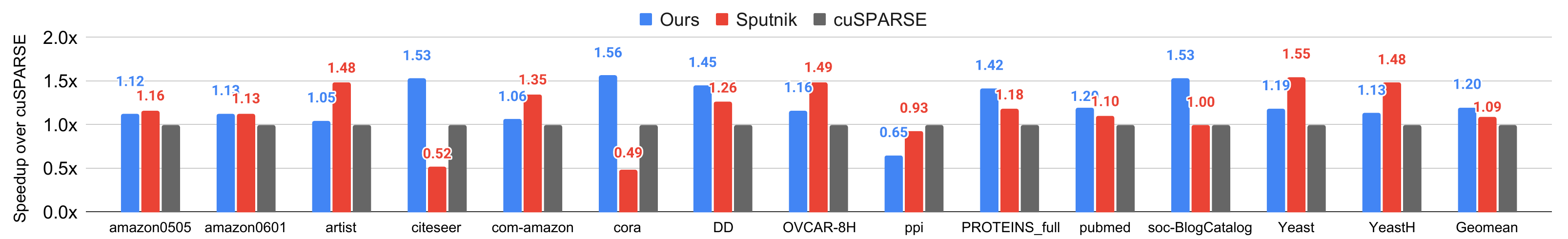}
  \caption{Comparison of our compiler-generated SpMM kernel, Sputnik and cuSPARSE on various sparse matrices.}
  \label{fig:spmm}
\end{figure*}
\subsection{Case Study: Structured SpMM}
\label{subsec:structured}

In this case study, we compare our compiler-generated block sparse matrix multiplication (SpMM) kernels with the state-of-the-art TorchBSR kernel~\cite{quansight2023bsr}, a hand-written Triton implementation integrated into PyTorch. TorchBSR requires 202 lines of Triton code, while our approach expresses in a single-line Einsum. We use the BlockGroupCOO format (Section~\ref{sec:format}) with $(32,32)$ block size and grouping along block rows. Figure~\ref{fig:bspmm} shows FP16 performance on $4096 \times 4096$ matrices, reported as speedup over dense matmul.

Our approach shifts the crossover point, where sparse outperforms dense, from about 40\% to 25\% sparsity. Our kernels consistently match or surpass TorchBSR, with a larger advantage at high sparsity. This is mainly because TorchBSR relies on the BCSR format, a block variant of CSR. Like CSR, BCSR incurs an $O(N)$ row-pointer overhead, where $N$ is the number of rows; even empty rows require storage and traversal. In hypersparse matrices, where most rows contain no nonzeros, this overhead becomes dominant~\cite{dcsr}. In contrast, BlockGroupCOO is COO-based and stores only nonzero blocks, eliminating per-row overhead and scaling more efficiently in the hypersparse regime.

\subsection{Case Study: Unstructured SpMM}

In this study, we compare our compiler-generated unstructured SpMM kernel with two state-of-the-art hand-written CUDA kernels: Sputnik~\cite{sputnik} and cuSPARSE~\cite{cusparse}. While Sputnik implements SpMM in approximately 2,000 lines of code and cuSPARSE's implementation details are unavailable due to NVIDIA's proprietary nature, our approach requires only a single Einsum expression (using GroupCOO format).

The operation we evaluate is $C_{m,n} = A_{m,k} * B_{k,n}$, where A is an unstructured sparse matrix. We selected real-world sparse matrices from TC-GNN datasets~\cite{tcgnn} with various sparsity patterns, including skewed distributions of nonzeros per row and matrices with large numbers of rows and nonzeros. For our experiments, we set the number of columns in C to 128.


Figure~\ref{fig:spmm} compares the performance of different kernels on FP32 inputs, reported as speedup relative to cuSPARSE. Our kernel achieves the highest average performance, running 1.2× faster than cuSPARSE, compared to 1.09× for Sputnik. For unstructured sparsity, however, no single kernel dominates across all test cases, since performance depends heavily on the distribution of nonzeros. Sputnik benefits from its row–permutation strategy that groups rows with similar nonzero counts, which gives it an advantage on datasets with highly skewed distributions (e.g., artist). 

We also evaluated FP16 inputs. These results are omitted from the figure because Sputnik provides only limited FP16 support. It can process matrices with fewer than $2^{16}$ rows, but cannot handle the large matrices in our test suite. In contrast, our kernel runs robustly across all benchmarks and, on average, performs 1.18× better than cuSPARSE in FP16.

Our kernel consistently delivers the best average performance. Importantly, it achieves this with a succinct, compiler-generated implementation, in contrast to Sputnik’s complex, hand-optimized code. Furthermore, our approach supports a broader range of problem specifications, including FP16, highlighting the effectiveness of compiler-based approach.

\begin{figure}[t]
  \centering

    \centering
    \includegraphics[width=\linewidth]{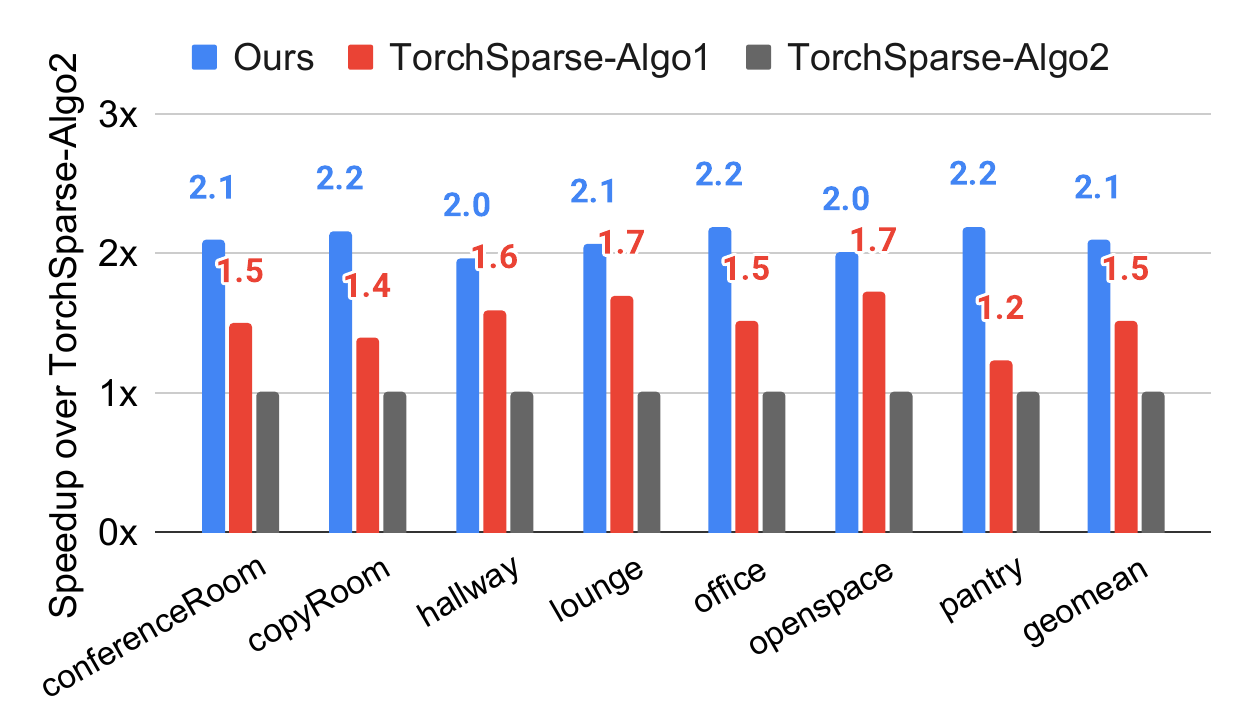}
    \caption{Speedup of our method vs. TorchSparse-Algo1 and TorchSparse-Algo2 for FP16 inputs.}
    \label{fig:spconvtf32}
  \label{fig:spconv}
\end{figure}

\subsection{Case Study: Point Cloud Sparse Convolution}

In this case study, we evaluate sparse convolution on point clouds and compare our generated kernel against TorchSparse \cite{torchsparse1}, a state-of-the-art library. TorchSparse implements sparse convolution using two different algorithms: (1) ImplicitGEMM \cite{torchsparse2} and (2) Fetch-on-Demand~\cite{torchsparse4}, each relying on distinct sparse formats and separate CUDA codebases. 

The point cloud sparse convolution can be expressed as:

\begin{center}
$Out_{x,m} = Map_{x,y,z} * In_{y,c} * Weight_{z,c,m}$ 
\end{center}

Here, $c$ and $m$ denote input and output channels, respectively. $In$ and $Weight$ are dense tensors and $Map$ is a 3D sparse tensor. If we store $Map$ in COO format, with its nonzero indices stored in $MAPX,MAPY,MAPZ$, and corresponding values in \texttt{MAPV}, the sparse convolution becomes:

\begin{center}
$Out_{MAPX_p,m} = MAPV_{p} * In_{MAPY_p,c} * Weight_{MAPZ_p,c,m}$
\end{center}

We can further optimize this by grouping entries by $MAPZ$, yielding the following einsum form:

\begin{center}
$Out_{MAPX_{p,q},m} = MAPV_{p,q} * In_{MAPY_{p,q},c} * Weight_{MAPZ_{p},c,m}$
\end{center}

This includes a batch matmul between $Out$, $In$, and $Weight$ (i.e., $pqm$ ← $pqc \times pcm$), enabling the use of Tensor Cores.

While TorchSparse requires over 4000 lines of CUDA code to implement and optimize these sparse convolution kernels across data types and channel sizes, our approach achieves the same computation using a single einsum expression.

Figure~\ref{fig:spconv} shows performance results for channel size 128 on seven real-world indoor point clouds from the S3DIS dataset (Area 6). We used a voxel size of 5 cm for quantization, as described in prior work~\cite{tacoucf}. Our method consistently outperforms both TorchSparse algorithms on FP16. This highlights the power of compiler-based approach compared to highly specialized handwritten implementations.

\subsection{Case Study: Equivariant Tensor Product}


\begin{table}[t]
\footnotesize
\centering
\begin{tabular}{c c c c c}
\Xhline{1pt}
\textbf{$\ell_{\max}$} & \textbf{Channels} & \textbf{Ours} & \textbf{cuequivariance} & \textbf{e3nn} \\
\Xhline{1pt}
\multirow{3}{*}{1} 
 & 16 & 8.3$\times$ & 2.6$\times$ & 1.0$\times$ \\
 & 32 & 4.2$\times$ & 1.5$\times$ & 1.0$\times$ \\
 & 64 & 2.3$\times$ & 0.9$\times$ & 1.0$\times$ \\
\hline
\multirow{3}{*}{2} 
 & 16 & 5.2$\times$ & 1.1$\times$ & 1.0$\times$ \\
 & 32 & 5.4$\times$ & 1.1$\times$ & 1.0$\times$ \\
 & 64 & 3.3$\times$ & 0.5$\times$ & 1.0$\times$ \\
\hline
\multirow{3}{*}{3} 
 & 16 & 2.6$\times$ & 0.5$\times$ & 1.0$\times$ \\
 & 32 & 3.6$\times$ & 0.6$\times$ & 1.0$\times$ \\
 & 64 & 2.5$\times$ & 0.3$\times$ & 1.0$\times$ \\
\Xhline{1pt}
\end{tabular}
\caption{Speedup of our method vs. cuequivariance and e3nn. Speedups are reported relative to e3nn (normalized performance). All experiments are conducted in FP32.}
\label{tbl:tp}

\end{table}

In this case study, we implement a fully connected tensor product in an equivariant neural network~\cite{e3nn} using our compiler. The fully connected tensor product—also referred to as the \texttt{uvw} mode—can be written as:

\begin{center}
$Z_{b,i,w} = CG_{i,j,k,l} * X_{b,j,u} * Y_{b,k} * W_{b,l,u,w}$
\end{center}

Here, $CG$ is a sparse 4D tensor, while all other tensors ($X,Y,W,Z$) are dense tensors. The indices $i,j,k,l$ in $CG$ determine the interaction pattern among the input tensors, while $u$ and $w$ denote input/output channels and $b$ denotes the batch dimension. We store $CG$ in COO format using index arrays $CGI$, $CGJ$, $CGK$, $CGL$ and corresponding nonzero values $CGV$, which allows us to rewrite the computation as:

\begin{center}
$Z_{b,CGI_p,w} = CGV_p * X_{b,CGJ_p,u} * Y_{b,CGK_p} * W_{b,CGL_p,u,w}$
\end{center}

To further optimize, we group entries by \texttt{CGL}, leading to:

\begin{center}
$Z_{b,CGI_{p,q},w} = CGV_{p,q} * X_{b,CGJ_{p,q},u} * Y_{b,CGK_{p,q}} * W_{b,CGL_p,u,w}$
\end{center}

This exposes a batched matrix multiply pattern between \texttt{Z}, \texttt{X}, and \texttt{W} of the form: $(bp)qw$ $\leftarrow$ $(bp)qu \times (bp)uw$. Here, the batch dimension is \texttt{bp} and the reduction dimension is \texttt{u}, allowing our compiler to apply Tensor Core acceleration.

We compare our grouped-COO einsum formulation against two hand-written libraries: \texttt{cuequivariance}\cite{cueq} and \texttt{e3nn}\cite{e3nn}. All experiments are conducted with a batch size of 10,000. To control the sparsity pattern of the \texttt{CG} tensor, we use real Clebsch-Gordan (CG) coefficients corresponding to different values of a hyperparameter $\ell_{\max}$. 

As shown in Table~\ref{tbl:tp}, while \texttt{cuequivariance} occasionally outperforms \texttt{e3nn}, our method consistently achieves higher speedups across all tested channel sizes and $\ell_{\max}$ values—up to $8.3\times$ faster and at least $2\times$ in every setting.

\subsection{Ablation Study} \label{subsection:analysis}


\begin{figure}[htbp]
    \centering
    \includegraphics[width=\linewidth]{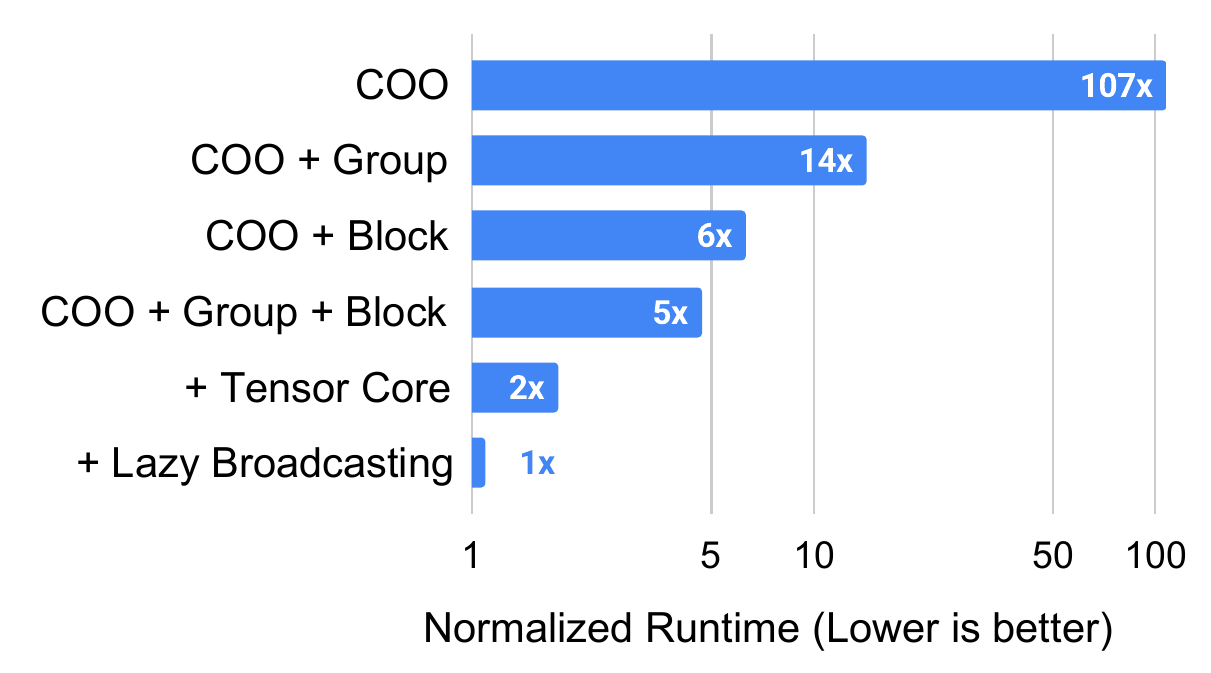}
    \caption{
Ablation study on structured SpMM. It shows how each optimization contributes to speedup over the baseline (COO without fusion). When all format and compiler optimizations are enabled, the performance surpasses that of the hand-written kernel (TorchBSR).
    }
    \label{fig:ablation}
\end{figure}

Figure~\ref{fig:ablation} shows an ablation study on structured SpMM to evaluate the impact of format selection and compiler optimizations. All matrices are of size 4096$\times$4096, with sparse matrix having 90\% uniform sparsity using 32$\times$32 dense blocks.

\textbf{Impact of Format:}  
We store the sparse matrix $A$ using three different formats. Each format is explained in Section~\ref{sec:format}:

\begin{itemize}[leftmargin=10pt]
    \item \textbf{COO:} A is stored in coordinate list (COO) format, where the row and column indices of nonzeros are stored as \texttt{AM} and \texttt{AK}. The indirect Einsum is:  $C_{AM_p,n} = AV_p * B_{AK_p,n}$.

    \item \textbf{COO + Group:} We group the row indices \texttt{AM} in groups of 16. 
    The indirect Einsum is:  $C_{AM_p,n} = AV_{p,q} * B_{AK_{p,q},n}$.

    \item \textbf{COO + Group + Block:} Once we block the matrix with 32$\times$32 dense block, we then group the block rows \texttt{AM} (by 4). The Einsum is:  $C_{AM_p,bm,n} = AV_{p,q,bm,bk} * B_{AK_{p,q},bk,n}$.
\end{itemize}

In the top three rows of Figure~\ref{fig:ablation}, we compare these formats without our compiler optimizations. In this setting, the PyTorch compiler separately launches gather, matrix multiplication, and scatter operations.




Grouping alone provides a substantial speedup, 8$\times$ over the baseline. This comes partly from reduced memory consumption: GroupCOO uses only 69\% of the memory required by COO. More importantly, grouping improves data reuse and reduces the number of scatter operations.

\begin{lstlisting}[basicstyle=\ttfamily\footnotesize]
# COO SpMM
parallel-for p in [0, NNZ):          
  for n in [0, N):          
    C[AM[p], n] += AV[p] * B[AK[p], n] # NNZ * N atomics

# GroupCOO SpMM (group size = g)
parallel-for p in [0, NNZ / g):      
  for n in [0, N):          
    acc = 0 # accumulate in register
    for q in [0, g): # nonzeros in a group       
      acc += AV[p, q] * B[AK[p, q], n]
    C[AM[p], n] += acc      # NNZ / g * N atomics
\end{lstlisting}

COO SpMM indirectly accesses \texttt{B} and \texttt{C} for every nonzero loop \texttt{p}, issuing an atomic update per element of \texttt{C}. In contrast, GroupCOO reuses the same row of \texttt{C} across all $g$ nonzeros in a group, accumulating their contributions in registers before issuing an atomic update. This improves locality for both \texttt{C} and \texttt{B}, and reduces the number of atomic calls by a factor of $g$, which explains much of the observed speedup.

Finally, we observe that combining grouping and blocking delivers even greater performance—up to 20× over the baseline. The blocked formats achieve this by enabling Tensor Core, while grouping improves data reuse.

\textbf{Impact of Compiler Optimizations:}  In the bottom two rows of Figure~\ref{fig:ablation}, we apply our compiler optimizations on top of the COO + Group + Block format. Enabling Tensor Core (TC) fusion delivers a 2.6× speedup over the default PyTorch compiler by fusing gather, matmul, and scatter into a single Triton kernel. This eliminates the need to materialize large intermediate tensors (which exceed 1.5GB when fusion is disabled) and avoids costly reloads from global memory. With fusion, the gathered data remains in shared memory and is fed directly into the Tensor Core without unnecessary trips to DRAM. Finally, Lazy Broadcasting further improves memory efficiency by eliminating the transposing and reshaping overhead in the Triton kernel, increasing the instruction level parallelism feeding into Tensor Cores.

\subsection{Comparison Against Other Sparse Compilers} \label{subsection:sparsecompiler}

\begin{table}[h]
\centering
\small
\begin{tabular}{lccc}
\toprule
Metric & \textbf{Ours} & \textbf{TACO} & \textbf{SparseTIR} \\
\midrule
Compile (s)       & 9.9   & \textbf{0.01}  & 0.32 \\
Autotune (s)         & \textbf{4.9}   & N/A (10 LoC) & N/A (860 LoC) \\
\makecell{FormatConvert (ms)} & 0.55 & \textbf{0.47} & 13.47 \\
Runtime (ms)           & \textbf{0.47} & 253.53 & 1.05 \\
\bottomrule
\end{tabular}
\caption{Performance, compilation cost, and conversion cost for the point cloud convolution on the \texttt{conferenceRoom} input with FP16 and channel size 128. TACO and SparseTIR lack autotuners, requiring users to manually search for the best schedule, often demands substantial time and code. }
\end{table}

While our compiler generates kernels that outperform hand-written kernels, two sources of overhead must be considered before kernel launch: (1) compilation and (2) format conversion. Focusing on these aspects, we compare \texttt{Insum} with other sparse GPU compilers, TACO~\cite{kjolstad_tensor_2017} and SparseTIR~\cite{ye2023sparsetir}, using the point cloud convolution example.

Compilation overhead typically consists of two parts: scheduling (tuning) to select tile sizes and format to store, followed by code generation and compilation. In deep learning workloads, this cost is easily amortized since compiled operators are cached and reused many times during inference. The difficulty arises because compilers lack an auto-scheduler, forcing developers to manually craft schedules. Using TACO, we spent hours identifying a schedule that still underutilized the GPU (no shared memory or Tensor Core usage), and SparseTIR required adopting a schedule of nearly 800 lines provided by its authors. In contrast, our compiler requires only a single indirect Einsum. We integrate the PyTorch compiler’s autotuning infrastructure to automatically discover efficient Triton configurations for our fused kernel, eliminating the need for manual schedule engineering. Although our compile+autotuning time is longer (14.8 seconds), the generated code is both the fastest and the only compiler-based approach to surpass the hand-written implementation (0.66 ms, TorchSparse). This process is fully automated, and its cost is easily amortized over repeated execution.

For the majority of applications, the format conversion overhead can also be amortized~\cite{won2023waco,ye2023sparsetir}. Many sparse workloads are static in structure, such as fixed graphs in GNNs~\cite{tcgnn} and pruned weight matrices~\cite{tacoucf}, allowing conversion to be done once offline and reused throughout inference. Even in dynamic scenarios such as point cloud convolution, the format conversion is done once and reused across multiple layers~\cite{torchsparse1}. Our framework shows moderate conversion cost (0.55 ms). The conversion kernel of SparseTIR is implemented on the CPU, resulting in significantly higher overhead (13.47 ms). TACO achieves the fastest conversion (0.47 ms), but its advantage is offset by prohibitively slow runtime performance. Thus, our approach strikes a balance: reasonable conversion time combined with the best kernel runtime.

\section{Conclusion}

This paper introduces a new approach for expressing sparse computations using indirect Einsums over fixed-length sparse formats, and presents Insum, a compiler that generates efficient GPU kernels by leveraging the PyTorch compiler. We further extend the PyTorch compiler to natively support Tensor Cores with lazy broadcasting, enabling Insum to generate fully fused gather–Einsum–scatter operations. Our evaluation demonstrates that sparse workloads can be expressed in a single indirect Einsum, while achieving better performance compared to hand-written sparse kernels.


\bibliographystyle{ACM-Reference-Format}
\bibliography{sample-base}

\end{document}